\documentclass{aa}
\usepackage{txfonts}
\usepackage{graphicx}
\usepackage{natbib}
\bibpunct{(}{)}{;}{a}{}{,}

\def\apjl{ApJ}%
\def\apjs{ApJS}%
\def\ao{Appl.~Opt.}%
\def\aap{A\&A}%
\begin{document}

\title{Are OPERA neutrinos faster than light because of non-inertial reference frames?}

\author{Claudio German\`a}
\institute{INAF-Astronomical Observatory of Padova, Italy\\ \email{claudio.germana@gmail.com}} 

\date{Received / Accepted}

\abstract
{Recent results from the OPERA experiment reported a neutrino beam traveling faster than light. The challenging experiment measured the neutrino time of flight (TOF) over a baseline from the CERN to the Gran Sasso site, concluding that the neutrino beam arrives $\sim 60$ ns earlier than a light ray would do. 
Because the result, if confirmed, has an enormous impact on science, it might be worth double-checking the time definitions with respect to the non-inertial system in which the neutrino travel time was measured. An observer with a clock measuring the proper time $\tau$ free of non-inertial effects is the one located at the Solar System Barycenter (SSB).}
{Potential problems in the OPERA data analysis connected with
the definition of the reference frame and time synchronization are emphasized. We aim to investigate the synchronization of non-inertial clocks on Earth by relating this time to the proper time of an inertial observer at SSB.}
{The Tempo2 software was used to time-stamp events observed on the geoid with respect to the SSB inertial observer time.}
{Neutrino results from OPERA might carry the fingerprint of non-inertial effects because they are timed by terrestrial clocks. The CERN-Gran Sasso clock synchronization is accomplished by applying corrections that depend on special and general relativistic time dilation effects at the clocks, depending on the position of the clocks in the solar system gravitational well. As a consequence, TOF distributions are centered on values shorter by tens of nanoseconds than expected, integrating over a period from April to December, longer if otherwise. It is worth remarking that the OPERA runs have always been carried out from April/May to November.}
{If the analysis by Tempo2 holds for the OPERA experiment, the excellent measurement by the OPERA collaboration will turn into a proof of the General Relativity theory in a weak field approximation. The analysis presented here is falsifiable because it predicts that performing the experiment from January to March/April, the neutrino beam will be detected to arrive $\sim$ 50 ns later than light.}


\titlerunning{}
\authorrunning{German\`a}

\maketitle

\section{Introduction}
The OPERA experiment \citep{2011arXiv1109.4897A} recently reported a neutrino beam traveling faster than light. The experiment measures the distributions of neutrino time emission/detection over a baseline from the CERN to the Gran Sasso (CNGS) site. Data are collected within runs lasting for several months\footnote{http://proj-cngs.web.cern.ch/proj-cngs/}.
In these data, the neutrino beam time of flight (TOF) turns out to be $\sim 60$ ns shorter than that calculated by taking the speed
 of light in vacuum. If confirmed, the result would have fundamental implications for modern physics. Basically, it reviews the fundamental postulate by Albert Einstein: ``Nothing can travel faster than light''\citep{1905AnP...322..891E}.\\
Although the implications are intriguing, one should consider possible systematic errors that are as yet undisclosed. The OPERA team pointed out that the research for possible sources of systematic errors is still in progress.

The scientific community has produced a plethora of works to interpret the puzzling neutrino speed result. These works were based on some systematic errors caused by the Earth's motion \citep{2011arXiv1110.3581M} or errors in synchronization of clocks by either a third clock moving into the gravitational field of Earth \citep{2011arXiv1109.6160C} or GPS satellites \citep{2011arXiv1110.2685V}, which were later reviewed \citep[][Armando V.D.B. Assis 2011\footnote{http://vixra.org/pdf/1110.0047v4.pdf}]{2011arXiv1110.2909B}. However, several works were based on superluminal neutrino and its implications, therefore considering the OPERA neutrino speed result free of systematic errors \citep[e.g.][]{2011arXiv1109.5445T,2011arXiv1111.4441L}.  

In this letter we focus on the question with respect to what a measure is made. The CNGS baseline has been measured through GPS benchmarks to better than 20 cm in the European Terrestrial Reference Frame (ETRF; Colosimo et al. 2011\footnote{http://operaweb.lngs.infn.it/Opera/publicnotes/note132.pdf}). However, from the point of view of a light ray traveling over CNGS (or a beam of particles), the TOF might need to be corrected for effects caused by the non-inertial observer ETRF, because the Earth rotates, orbits around the Sun and exhibits other motions (precession, nutation, polar motion).
\mbox{Nanosecond-precision} measurements of OPERA are demanded to be in pulsar timing. To study the actual pulsar clock behavior, the time tagged by clocks running on Earth has to be corrected to that by an inertial observer. 

A reference frame that approximates an inertial one is the Barycentric Celestial Reference System (BCRS), which is non-rotating and located at the Solar System Barycenter (SSB). To turn the problem into the inertial frame BCRS, we use the Tempo2 software \citep{2006MNRAS.369..655H, 2006MNRAS.372.1549E}, a software conceived to correct for non-inertial effects the time of arrival of photons coming from pulsars. This uses the most accurate planetary ephemerides to date (Standish 1998\footnote{ftp://ssd.jpl.nasa.gov/pub/eph/planets/ioms/de405.iom.pdf}), in agreement with the IAU 2000 resolutions. 

\section{From ETRF to BCRS: Tempo2}
Tempo2 is the software that best-models light rays traveling from a pulsar to the observatory, with an accuracy of $\sim1$ ns. The software corrects the time of arrival of photons, as measured at the observatory, for general relativistic effects on both the photon-path and clocks running in a gravitational field, against which time is measured. 
In this framework Tempo2 is used to study how the inertial observer at SSB would see a signal traveling over CNGS. 

After transforming the ETRF coordinates of the sites into the International Terrestrial Reference System ITRS\footnote{For full accuracy it is suggested to use coordinates in ITRS. ETRF coordinates were transformed into ITRS through the online converter at  http://www.epncb.oma.be/$\_$dataproducts/coord$\_$trans/index.php.}, they are inserted into Tempo2. The software calculates the ephemerides of the  vector components \textbf{r}($\Sigma,t_{UTC}$), pointing from the SSB to the site $\Sigma$ on the geoid at terrestrial UTC time $t_{UTC}$. 
The software prints the components of the vector \textbf{r}$'$($\Gamma,t_{UTC}$) from the SSB to the geocenter $\Gamma$ and those of the vector \textbf{s}($\Sigma,t_{UTC}$) from the geocenter to the site. The vector \textbf{r}$'$($\Gamma,t_{UTC}$) is constructed in the barycentric frame BCRS, while \textbf{s}($\Sigma,t$) refers to the Geocentric Celestial Reference System (GCRS), which is non-rotating and located at geocenter. However,  \textbf{s}($\Sigma,t_{UTC}$) would have been equivalent if it had been constructed in the frame BCRS \citep{2006MNRAS.372.1549E}. In \textbf{s}($\Sigma,t_{UTC}$) the software accounts for Earth rotation, precession-nutation, polar motion and irregularities in both polar motion and rotation.\\
The vector \textbf{r}($\Sigma,t_{UTC}$) in BCRS reads
\begin{equation}
\mathbf{r}(\Sigma,t_{UTC})=r_{x}(\Sigma,t_{UTC})\mathbf{u}_{x}+r_{y}(\Sigma,t_{UTC})\mathbf{u}_{y}+r_{z}(\Sigma,t_{UTC})\mathbf{u}_{z},
\end{equation}
where $r_{i}(\Sigma,t_{UTC})=r'_{i}(\Gamma,t_{UTC})+s_{i}(\Sigma,t_{UTC})$ ($i=x, y, z$) and \textbf{u}$_{x}$, \textbf{u}$_{y}$, \textbf{u}$_{z}$ are unit vectors in the barycentric frame BCRS.

If the light ray is emitted at the CERN at the space-time coordinate \textbf{r}$(CN,t_{UTC})$ and is detected at Gran Sasso at \textbf{r}$(GS,t'_{UTC})$, the two events are linked by the equation
\begin{equation}\label{eq1}
c\left(\tau'\left(GS,t'_{UTC}\right)-\tau\left(CN,t_{UTC}\right)\right)=\left|\mathbf{r}\left(GS,t'_{UTC}\right)-\mathbf{r}\left(CN,t_{UTC}\right)\right|,
\end{equation}
where $c$ is the speed of light, $\tau\left(\Sigma,t_{UTC}\right)$ is the Barycentric Coordinate Time (TCB) as a function of the site $\Sigma$ and UTC time. The TCB is the proper time experienced by the inertial observer at the SSB \citep{1998A&A...336..381S}. The TCB time is linked to UTC time by the relation
\begin{equation}\label{eq02}
\tau\left(\Sigma,t_{UTC}\right)=t_{UTC}+ls+\Delta_{E}\left(\Sigma,t_{UTC}\right),
\end{equation}
where $ls$ are leap seconds to tie the UTC time to Terrestrial Time (TT) \citep[$\sim 66$ s;][]{1992A&A...265..833S}, $\Delta_{E}\left(\Sigma,t_{UTC}\right)$ is the Einstein delay, which accounts for relativistic effects on clocks running in a gravitational field.\\
Expressing equation (\ref{eq1}) in UTC we obtain
\begin{eqnarray}\label{eq01}
c\left(t'_{UTC}-t_{UTC}\right)&=&\left|\mathbf{r}\left(GS,t'_{UTC}\right)-\mathbf{r}\left(CN,t_{UTC}\right)\right|+{}\nonumber\\
&{}-&c\left(\Delta'_{E}\left(GS,t'_{UTC}\right)-\Delta_{E}\left(CN,t_{UTC}\right)\right),
\end{eqnarray}
from which the TOF ($t'_{UTC}-t_{UTC}$) of the signal in UTC \emph{proper time} can be deduced.\\
The term $\Delta'_{E}(GS,t'_{UTC})-\Delta_{E}(CN,t_{UTC})$ is the difference of the Einstein delay at the two clocks and at $t_{UTC}'$, $t_{UTC}$. 
In the OPERA experiment clocks were not corrected for $\Delta_{E}$, therefore they tick at UTC time, which is not a proper time. 

\section{Results}
\begin{figure}[!t!]
\resizebox{\hsize}{!}{\includegraphics[width=0.3\textwidth]{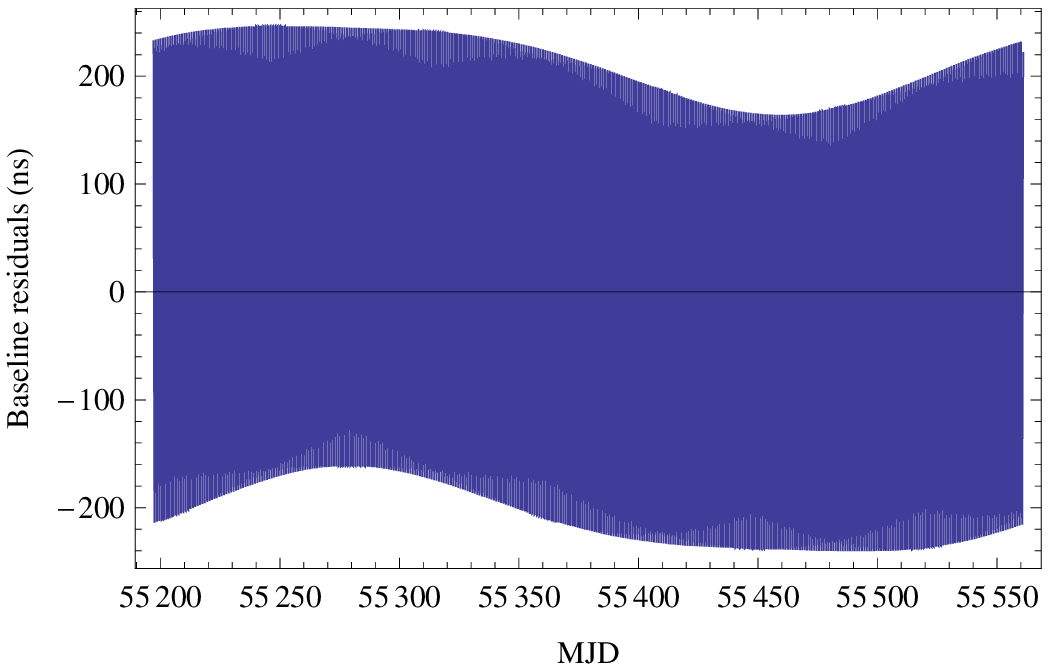}}
\resizebox{\hsize}{!}{\includegraphics[width=0.3\textwidth]{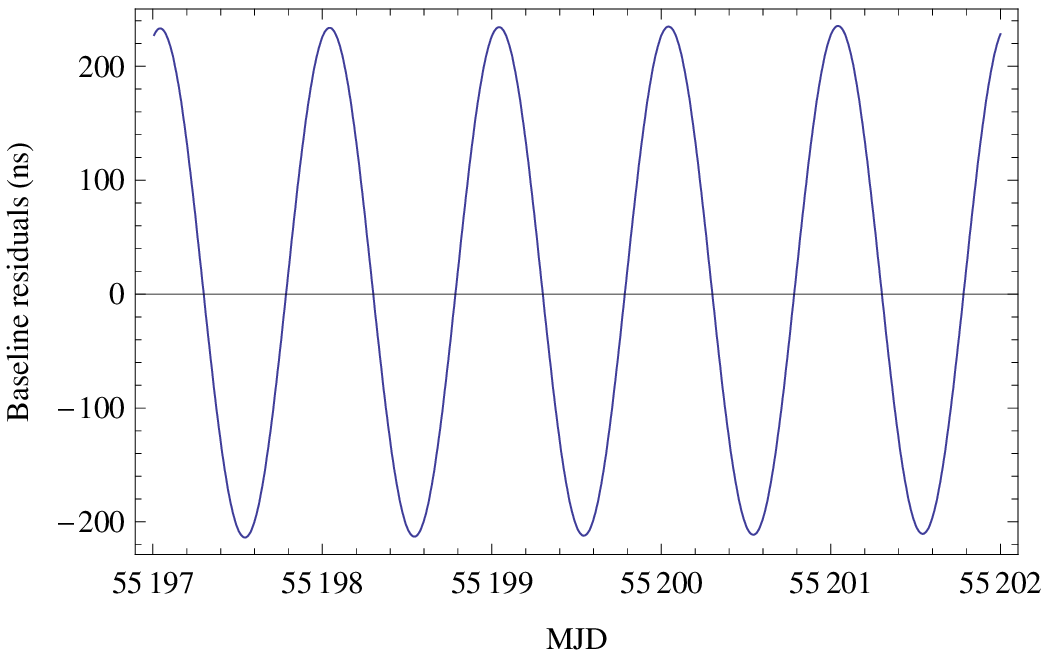}}
\caption{\emph{Top}: Timing residuals after subtracting the expected TOF from that of Eq. (\ref{eq01}), neglecting the Einstein delay term. The simulation is performed for the year 2010, expressed in Modified Julian Day (MJD=JD-2400000.5). The bin time of the temporal series is 10 minutes. \emph{Bottom}: Zoom-in over five days.}\label{fig1}.
\end{figure}
The geocentric coordinates of both the CERN and the Gran Sasso sites reported in Colosimo et al. (2011) were inserted into Tempo2. These coordinates refer to the origin of the OPERA detector reference frame at Gran Sasso and to the target focal point at the CERN, 730534.61 m away. An additional baseline is that between the beam current transformer (BCT) and the focal point at the CERN \citep[743.391 m;][]{2011arXiv1109.4897A}. Because the geocentric coordinates of the BCT were not found in the literature, we assume the CNGS baseline to be 730534.61 m. An additional baseline of 743.391 m is equal to \mbox{$\sim$ 20 arcsec} on the Earth's surface and would induce a negligible timing correction in the relativistic term $\textbf{v}\cdot\textbf{s}/c^{2}$  (see Sec. \ref{sec31}). 

\subsection{Tying non-inertial clocks on Earth to the clock at BCRS}\label{sec31}
We now study the TOF from equation (\ref{eq01}) neglecting the term $\Delta'_{E}(GS,t'_{UTC})-\Delta_{E}(CN,t_{UTC})$, like in OPERA. Fig.~\ref{fig1} (top) shows the TOF of the signal after subtracting the value one expects, 730534.61 m/$c$=0.002436801 s. There is an excess of up to $\sim240$ ns modulated by both an annual component and the Earth's rotation (Fig.~\ref{fig1} bottom). Fig.~\ref{fig1} shows that the neutrino TOF is never constant and depends on the epoch at which the measurement is performed.

\begin{figure}[!t!]
\resizebox{\hsize}{!}{\includegraphics[width=0.3\textwidth]{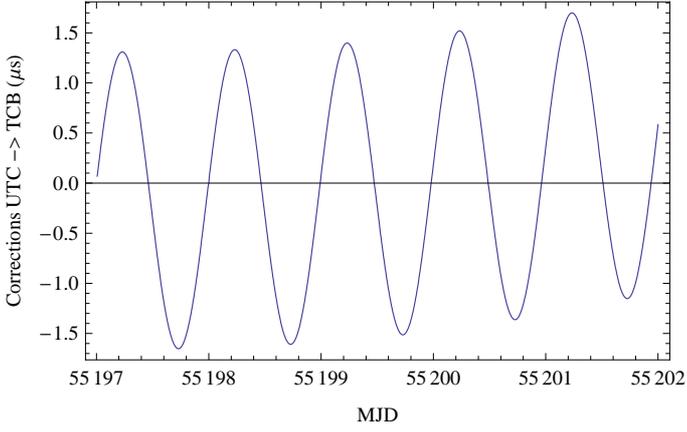}}
\caption{Einstein delay correction to tie the clock at CERN from UTC to TCB time after removing the fit first-order polynomial.}\label{fig2}
\end{figure}

To understand what happens, we need to analyze the Einstein delay difference $\Delta'_{E}(GS,t'_{UTC})-\Delta_{E}(CN,t_{UTC})$. We print the Einstein delay for the clock at the CERN to tie UTC to TCB time (eq. [\ref{eq02}]). 
The amount of the correction at MJD=55197.0 (January 1, 2010) is up to $\sim16$ s and grows with a slope of \mbox{$\sim1.5\times10^{-8}$ s/s} since MJD=43144.0003725. 
The linear drift of \mbox{$\sim1.5\times10^{-8}$ s/s} takes into account the linear term of the Einstein time-dilation integral, and another term accounting for the gravitational redshift due to Earth's potential \citep{1999A&A...348..642I}. The Einstein integral accounts for the special relativistic time-dilation ($v^{2}/c^{2}$, $v$ velocity of the geocenter) and the gravitational redshift at geocenter ($U/c^{2}$, $U$ gravitational potential where the geocenter moves). Because $v\sim30$ km/s, the special time-dilation is $v^{2}/c^{2}\sim10^{-8}$ s/s.\\ 
Fig.~\ref{fig2} shows the behavior of the Einstein delay correction after fitting a first-order polynomial over five days and subtracting it.\footnote{The coefficient of the polynomial obtained from the fit is \mbox{$\sim1.5\times10^{-8}$ s/s}, as expected.} The overall linear behavior hides a diurnal modulation of some microseconds owing to Earth's rotation. The modulation originates from the special relativistic time-dilation on the geoid with respect to the geocenter. The correction reads $\textbf{v}\cdot\textbf{s}/c^{2}$, with $\textbf{v}$ the velocity vector of the geocenter and $\textbf{s}$ the site position vector pointing from the geocenter to the site. Because $\left|v\right|\sim30$ km/s and $\left|s\right|\sim6300$ km (radius of Earth), the corrections amount to up to $\sim$ 2 $\mu$s modulated by Earth's rotation.\\ 
The non-zero slope in Fig.~\ref{fig2} after removing the linear drift signifies that the corrections hide other terms. The Einstein delay integral, referring to the motion of the geocenter, has several periodic terms \citep{1992A&A...265..833S}. 

\begin{figure}[!t!]
\resizebox{\hsize}{!}{\includegraphics[width=0.3\textwidth]{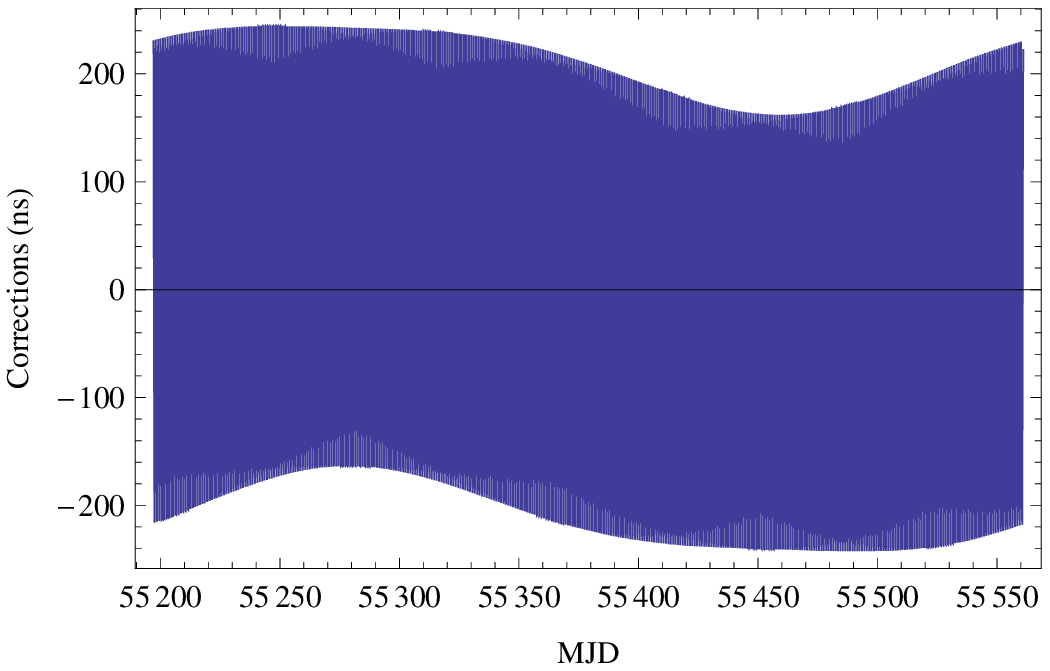}}
\resizebox{\hsize}{!}{\includegraphics[width=0.3\textwidth]{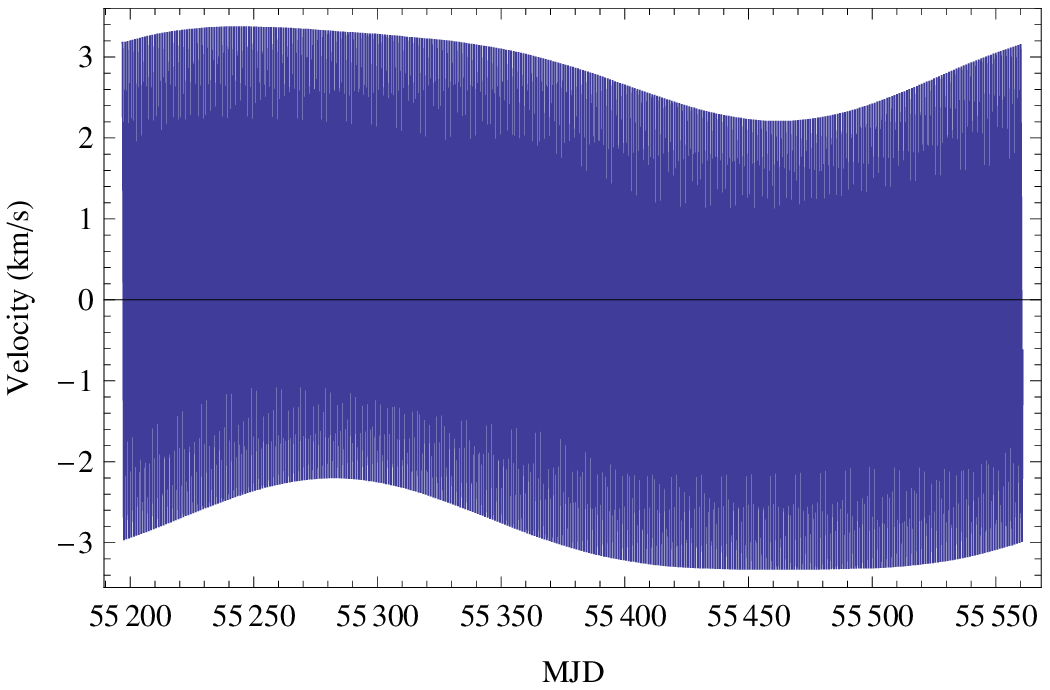}}
\caption{\emph{Top}: Corrections that the non-inertial observer ETRF should apply to keep the clock at Gran Sasso synchronized with that at the CERN. \emph{Bottom}: The difference behavior of the velocity components parallel to the position vector \textbf{s} at the two sites for the year 2010.}\label{fig3}.
\end{figure}

The inertial observer ties the UTC time of the clock at Gran Sasso to the TCB as well. 
The corrections to the clock at the CERN are subtracted from the Gran Sasso clock corrections to give the Einstein delay difference $\Delta'_{E}(GS,t'_{UTC})-\Delta_{E}(CN,t_{UTC})$. Fig.~\ref{fig3} (top) shows the result. The behavior is equal to that seen in Fig.~\ref{fig1}. 
Hence, the cause for the residual on each single TOF measurement shown in Fig.~\ref{fig1} is that the UTC time is not corrected for relativistic effects through the Einstein delay difference $\Delta'_{E}-\Delta_{E}$, as equation (\ref{eq01}) would instead require. Correcting the UTC time for the Einstein delay implies null residuals in Fig.~\ref{fig1}, as one would expect.\\
Because the UTC time in OPERA is not corrected for \mbox{$\Delta'_{E}-\Delta_{E}$}, the two clocks at the CERN and Gran Sasso are not properly synchronized. To keep them synchronized, the non-inertial observer at ETRF has to apply the corrections in Fig.~\ref{fig3} (top) to the UTC time at Gran Sasso.

Like Fig.~\ref{fig1} (bottom), the corrections to keep clocks synchronized oscillate with a diurnal behavior. Consequently, events that are simultaneous in TCB time are simultaneous in UTC time as well only when the correction in Fig.~\ref{fig3} (top) is zero. This happens whenever the special relativistic correction on the geoid with respect to the geocenter, $\textbf{v}\cdot\textbf{s}/c^{2}$, is equal at both sites.\\ To verify that the behavior in Fig.~\ref{fig3} (top)  is indeed caused by differences in the term $\textbf{v}\cdot\textbf{s}/c^{2}$ between the sites, Fig.~\ref{fig3} (bottom) shows the difference of the dot product $\textbf{v}\cdot\textbf{j}$ between the CERN and Gran Sasso (\textbf{j} unit vector along \textbf{s}). The difference of the velocity component parallel to \textbf{s} is up to 3 km/s, modulated by both diurnal and annual components. Its shape reflects that of the  relativistic corrections in Fig.~\ref{fig3} (top). A difference in velocity of up to 3 km/s corresponds to a difference in timing corrections $(\delta\left|v\right|\times\left|s\right|)/c^{2}\sim 210$ ns. The annual component comes from the modulation over the seasons of the angle between $\textbf{v}$ and $\textbf{s}$ \citep[see also][]{2011arXiv1110.3581M}.

\subsection{Distribution of timing corrections}
\begin{figure}[!t!]
\resizebox{\hsize}{!}{\includegraphics[width=0.3\textwidth]{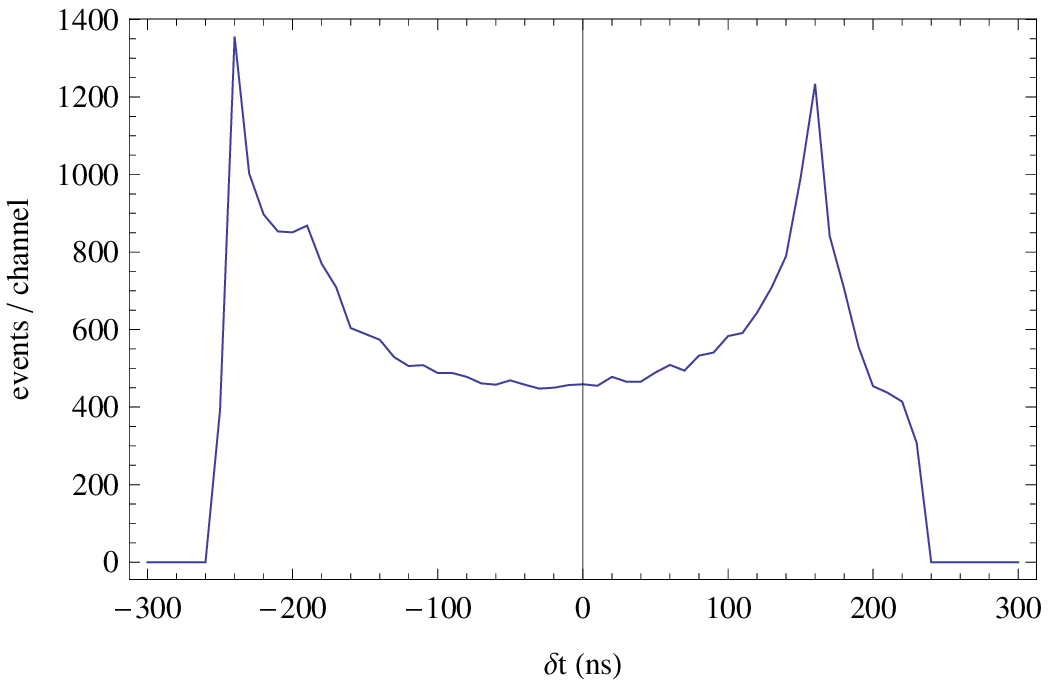}}
\resizebox{\hsize}{!}{\includegraphics[width=0.3\textwidth]{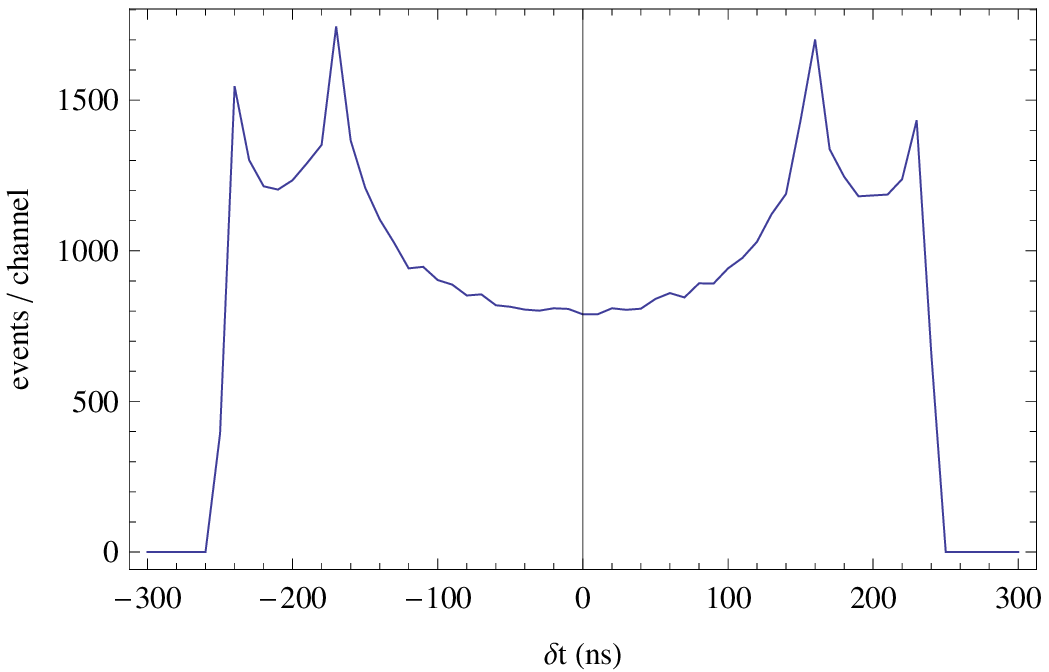}}
\resizebox{\hsize}{!}{\includegraphics[width=0.4\textwidth]{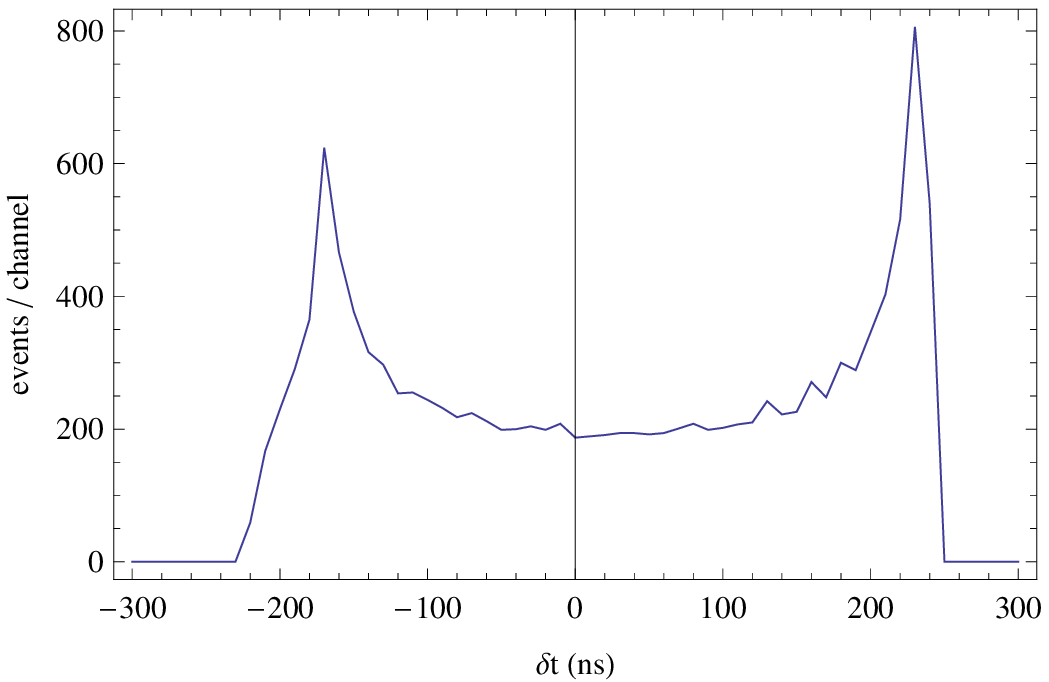}}
\caption{Distribution of timing corrections. \emph{Top}: Distribution calculated over a typical OPERA run (April to November 2010). It would imply an overall \mbox{miss-synchronization} of $\sim$ -80 ns. \emph{Middle}: Distribution over an entire year. \emph{Bottom}: Distribution from January 1 to March 31. It would imply an overall \mbox{miss-synchronization} of $\sim$ +50 ns.}\label{fig5}.
\end{figure}

To check whether the overall effect of the corrections in Fig.~\ref{fig3} completely vanishes, we can study their distribution.\\ We divide the y-axis of Fig.~\ref{fig3} (top) into 10 ns long channels. For each channel we count the number of events, integrating over a typical OPERA run. Fig.~\ref{fig5} (top) shows the distribution over an OPERA run from April 29 to November 22, 2010 (from MJD=55314.0 to MJD=55522.0). It peaks at both the \mbox{-240} ns and +160 ns channel, therefore it might induce an overall error on the clock synchronization of $\sim$ -80 ns\footnote{The $\sim$ -80 ns overall effect holds for the last OPERA results release as well \citep[][]{2011arXiv1109.4897A}, because the run lasted from October 21 to November 7, 2011.}.\\ 
To investigate farther out, we study the distribution for measures taken over an entire year, from January 1 to December 31, 2010. Fig.~\ref{fig5} (middle) shows that the distribution is symmetric with respect to channel zero, inducing a null effect on the clock synchronization. 

The difference between the two distributions in Fig.~\ref{fig5} (top-middle) might be explained in terms of the variation of the angle between $\textbf{v}$ and \textbf{s} over the seasons (see also Fig.~\ref{fig3} bottom). 
The OPERA runs have always been carried out from April/May to November, i.e., they missed the period from January to March that would be needed to complete one angle cycle. Therefore, if the measures are taken only in a window of the entire Earth orbit, it might be possible that the clock synchronization is affected by this systematic error and the non-inertial observer on Earth would measure a TOF shorter by \mbox{$\sim$ 80 ns}. Because he is unaware of non-inertial effects, the non-inertial observer should take care to carefully synchronize the clocks.  Alternatively, the measurements could be integrated over one angle cycle: In this way, from the perihelion to the aphelion and back, the effect vanishes.\\
If we add the $\sim$ +80 ns correction to the discrepancy claimed in the OPERA experiment (-60 ns), the result might change: We obtain a neutrino beam arriving later than light.    

\section{Conclusions}
The analysis by Tempo2 suggests that clocks on Earth should be tied to the TCB time to ensure their synchronization. The TCB time is the proper time experienced by an inertial observer located at the solar system barycenter. The UTC time ensures synchronization only when the relativistic correction $\textbf{v}\cdot\textbf{s}/c^{2}$ \citep[dubbed differential special time-dilation;][]{2006MNRAS.372.1549E} is equal at both clocks. 
Differences in this correction of up to $\sim240$ ns are seen and are modulated by both the Earth's rotation and revolution.  

The overall effect of a non-synchronization of clocks may vanish over one Earth orbit. Integrating the non-synchronized clock over months might induce an overall effect on the order of several tens of nanoseconds.\\
If the analysis described in this letter holds for the OPERA experiment, the excellent measurement by \citet{2011arXiv1109.4897A} would turn into a proof of relativistic theories \citep{1905AnP...322..891E,1916AnP...354..769E}. The analysis might cast doubts on interpretations dealing with superluminal neutrino that are, or are not, accounting for relativistic effects \citep[e.g.][]{2011arXiv1109.5445T,2011arXiv1111.4441L}.

To check whether or not the present study holds, it might be interesting to see results from OPERA runs performed in a different period of the year than usual.
Fig.~\ref{fig5} (bottom) shows the distribution of the corrections for non-inertial effects from January-March. Again, it is not symmetric with respect to channel zero, but shows an opposite result to that of the distribution on the top. From this result the neutrino beam should arrive $\sim$ 50 ns later than light.   

\footnotesize{\section*{Acknowledgments}
I would like to thank Andrej \v{C}ade\v{z} (Faculty of Mathematics and Physics, University of Ljubljana) and Massimo Calvani (INAF-Astronomical Observatory of Padova) for stimulating discussions. I also thank Rodolfo Angeloni (Departamento de Astronom\'ia y Astrof\'isica, Pontificia Universidad Cat\'olica de Chile) for his kind support.}

\bibliographystyle{aa}
\bibliography{biblio}

\end{document}